\documentclass[english]{emulateapj}
\usepackage[T1]{fontenc}
\usepackage[latin9]{inputenc}
\usepackage{verbatim}
\usepackage{amssymb}
\usepackage{color}
\begin{document}

\title{Effects of Dynamical Evolution of Giant Planets on Survival of
Terrestrial Planets}

\author{Soko Matsumura\altaffilmark{1}}
\author{Shigeru Ida\altaffilmark{2}}
\author{Makiko Nagasawa\altaffilmark{2}}

\altaffiltext{1}{Department of Astronomy and Astrophysics,
University of Maryland, College Park, MD 20741.}

\altaffiltext{2}{Department of Earth and Planetary Sciences, Tokyo Institute of
Technology, Tokyo, Japan.}

\begin{abstract}
The orbital distributions of currently observed extrasolar giant planets allow marginally 
stable orbits for hypothetical, terrestrial planets.   
In this paper, we propose that many of these systems may not have additional planets on these ``stable'' orbits, since 
past dynamical instability among giant planets could have removed them. 
We numerically investigate the effects of early evolution of multiple giant planets 
on the orbital stability of the inner, sub-Neptune-like planets which are modeled as test particles, and determine their dynamically unstable region.
%
Previous studies have shown that the majority of such test particles are ejected out of the system as a result of close encounters with giant planets.
Here, we show that secular perturbations from giant planets can remove test particles at least down to 10 times smaller than their minimum pericenter distance. 
Our results indicate that, unless the dynamical instability among giant planets is either absent or quiet like planet-planet collisions, most test particles down to $\sim0.1\,$AU within the orbits of giant planets at a few AU may be gone.
In fact, out of $\sim30\,\%$ of survived test particles, about three quarters belong to the planet-planet collision cases.
We find a good agreement between our numerical results and the secular theory, and present a semi-analytical formula which estimates the dynamically unstable region of the test particles just from the evolution of giant planets.
Finally, our numerical results agree well with the observations, and also predict the existence of hot rocky planets in eccentric giant planet systems.
\end{abstract}


%
\section{Introduction}\label{intro}
The solar system has a rich planetary architecture with four inner terrestrial planets and four outer giant planets. 
Whether such a planetary system is common or not has been a long-standing question in astronomy.
The recent observations started revealing that multiple-planet systems are common, especially 
among close-in sub-Neptune-like planets \citep[e.g.,][]{Mayor11ap,Batalha12ap}.
However, we are yet to find systems similar to ours.
Future observations would be able to tell us whether some of these systems with multiple sub-Neptune-like planets 
are accompanied by giant planets on further orbits, and/or known giant-planet systems have terrestrial planets interior to their orbits. 

In the meantime, we can explore whether the currently observed planetary systems allow the existence of additional planets.
Such an attempt has been made by investigating the orbital stability of hypothetical terrestrial planets in the observed giant-planet systems \citep[e.g.][]{Gehman96,Jones01,Noble02,Menou03,Asghari04}.
These studies showed that some of the observed systems could indeed host dynamically stable Earth-like planets in their habitable zones (HZs).
\cite{Menou03} modeled terrestrial planets as test particles, and investigated dynamical habitability of these planets in 85 then-known exoplanetary systems.
They found that about $25\%$ of their samples could host habitable, Earth-like planets.
These studies, however, do not take account of the effects of potential early evolution of giant-planet systems on formation and/or survival of terrestrial planets.

The migration and dynamical instability of giant planets are two major threats to formation and stability of Earth-like planets interior to their orbits. 
\cite{Armitage03} pointed out the potentially hazardous effect of giant planet migration on formation of terrestrials, and suggested that these planets preferentially exist in systems where massive giants did not migrate significantly.
Various numerical simulations, however, are generally agreed that terrestrial planet formation is not necessarily prevented by giant planet migration, when eccentricity excitation timescales for (proto-)terrestrial planets are long compared to migration timescales of giant planets \citep[e.g.,][]{Mandell03,Lufkin06,Raymond06}.
These less massive planets can survive by either being shepherded interior to the migrating giant planet, or scattered exterior to its orbit \citep[e.g.,][]{Zhou05,Fogg05}.
However, the survival rate of less massive planets strongly depends on the migration speed of the massive one \citep[e.g.,][]{Ward95,Tanaka99,Edgar04}. 
\cite{Mandell03} showed that the survival percentage of planets decreases from $40\%$ to $15\%$ when the migration timescale for a giant planet increases from 0.5 to $2\,$Myr.

How the dynamical instability of giant planets affects terrestrial planet formation have also been investigated \citep[e.g.,][]{Thebault02,Levison03,Veras05,Veras06,Raymond11,Raymond12}.
\cite{Veras05} showed that, when a moderately eccentric giant planet is left at a few AU, over $95\%$ of test particles are {\it ejected} 
--- i.e., terrestrial planet formation is more difficult in such a system.
However, their study focused on the stability of planets in the HZs ($0.75-1.25\,$AU). 
Although their follow-up study has extended the orbital radii down to $0.5\,$AU, the recent observations have found many sub-Neptune-sized planets within the orbital radius of Mercury (about $0.4\,$AU). 
Thus, we must take account of much closer-in planets to discuss the effects of dynamical instability among giant planets on the survival of these short-period, low-mass planets. 

In this paper, we study the orbital evolution of terrestrial planets when giant planets become dynamically unstable, and show that terrestrial planets far away from giants can also be removed.  
The numerical methods and initial conditions of our simulations are described in Section 2.  In all of our simulations, we approximate terrestrial planets as test particles, and follow their evolution as three Jupiter-mass planets become dynamically unstable. 
As we show in Section 3, many of these planets merge with the central stars, rather than being ejected out of the systems. 
In Section~\ref{results_general}, we compare the numerical results with the secular theory, and show that mergers are largely due to secular interactions.  
In Section~\ref{results_stability}, we present a convenient formula which estimates the dynamically unstable region of terrestrial planets from the orbital evolution of a giant-planet system alone. 
In Section~\ref{results_Joutcomes}, we discuss how the fates of test particles depend on the dynamical outcomes of giant planets. 
In Section~\ref{results_freshtps}, we investigate the importance of past evolution of giant planets on the orbital stability of test particles.
In Section~\ref{results_obs}, we compare our results with the distributions of observed exoplanetary systems.
Finally, we address the directions of future studies, and summarize our work in Section 4.
\section{Method}\label{method}
We perform numerical simulations of planetary systems by using a publicly available N-body code {\it Mercury} \citep{Chambers99}.
For all the simulations presented in this paper, we use the Bulirsch-Stoer method with a tolerance parameter of $10^{-12}$ to ensure the high accuracy.
We run simulations up to 10\,Myr, unless it is noted otherwise.  
In the code, a planet-star merger occurs when the pericenter distance becomes less than the stellar radius, while a planet-planet collision occurs when the distance between them becomes less than the sum of their radii. 
Furthermore, the planets that go beyond 1000\,AU are assumed to be ejected.

The standard planet formation theory predicts that gas giant planets are formed beyond the ice line.  
However, there is no consensus about how many such planets being formed per a typical protostellar disk.  
Since we are interested in the effects of dynamical instability among giant planets on survival or formation of terrestrial planets, 
we follow the approach of \cite{Veras05} and use three giant-planet systems.   
This is also the smallest number of planets for which there is no analytical stability limit \citep[e.g.,][]{Marchal82,Gladman93,Chambers96}.
Our initial conditions assume that giant planets initially orbit beyond the ice line, and that the dynamical instability occurs after the gas disk has dissipated.
The dynamical instability among giant planets in such systems would not be able to recover 
the orbital distribution of currently observed exoplanets within $\sim1\,$AU 
\citep[e.g.,][]{Adams03,Chatterjee08}.
The formation of closer-in planets require either planet migration \citep[e.g.,][]{Lin96,Adams03,Moorhead08,Matsumura09} or dynamical evolution of giants combined with tidal interactions with the central star 
\citep[e.g.][]{Fabrycky07,Wu07,Nagasawa08,Naoz11,Wu11}. 
However, recent radial velocity observations show that there is a jump in the number of single-planet systems beyond $\sim1\,$AU \citep{Wright09}.
Thus, the systems we consider here and their dynamical outcome may not be so unrealistic. 

We test three sets of initial conditions (Set\,0, 1, and 2). 
For the default set (Set\,0), we assume that all three giant planets have a Jupiter mass, and the initial semimajor axes of 3.8, 5, and 6.5 AU, respectively.  
We also assume that initial eccentricities of Jupiters are zero, and they have non-zero, 
but small initial inclinations of 0.001, 0.005, and 0.003 degrees, respectively.
The terrestrial planets are modeled as test particles in this work, 
which is a reasonable assumption since the mass ratio of Earth to Jupiter is small ($M_E/M_J\sim 3\times10^{-3}$).   
We consider 11 test particles that are distributed over $0.1-3\,$AU, and assume that all of them initially have circular and coplanar orbits.  
Each test particle has a randomly assigned orbital phase, but their distribution is common for all the simulations. 
We perform 40 scattering simulations, where we only change the initial orientations of giant planets' orbits.  

The instability among equal-mass planets tends to result in the high eccentricities of survived planets compared to unequal-mass systems \citep[e.g.,][]{Ford08}.  
Thus, we do not intend to match the eccentricity distribution of giant planets with the observed one.  
However, since we are only interested in the range of the stable region, and not the rate of stable test particles, this assumption would not change our conclusion. 
To assess the effect of a mass variation, we also perform Set\,1, where we keep the initial orbital distribution of giant planets the same, and change the mass of one of the giants to $0.5M_J$ instead of $1M_J$.  By varying the initial location of the smaller giant from innermost, middle, and outermost, we perform 42 runs in total for Set\,1.   

The closely-separated orbital distribution in a default set generally leads to a dynamical instability within a short period of time.
Naturally, the actual planet formation is unlikely to lead to such quasi-unstable systems.  
It's expected that, once the instability occurs, the final outcome of dynamical evolution of planets will not be largely affected by the initial distribution of planets, as long as the total energy and the angular momentum do not differ much \citep[e.g.,][]{Juric08,Nagasawa08}.
To test this point, we perform Set\,2, where we keep the giant planetary mass the same, and change the orbital separations, eccentricities, and inclinations.  We choose the wider orbital separation so that the time to dynamical instability would be $\sim1\,$Gyr if giant planets were on nearly circular orbits.  From \cite{Chatterjee08}, we estimate such orbital radii are 3.8, 6, and 9.3 AU, where orbits are separated by $K\sim5$ mutual Hill radii. To speed up the simulations, we also choose higher eccentricities (0.15, 0.2, and 0.1 for the innermost, middle, and outermost planet, respectively) and inclinations (4.3, 5.7, and 2.9 degrees).
We run 40 such simulations for Set\,2.

Since all of these sets give similar results to one another, we focus our attention on the default set in Section~\ref{results}, and discuss the other sets in Section~\ref{summary}.
\section{Results}\label{results}
Here, we discuss our default set (Set\,0).
In Section~\ref{results_general}, we present the overall outcomes of our simulations.  
We find a good agreement between our results and the expectations from the secular theory. 
In Section~\ref{results_stability}, we propose a formula which can predict the dynamically unstable region of test particles.
In Section~\ref{results_Joutcomes}, we discuss the dependence of fates of test particles on dynamical outcomes of giant planets.
We show that the dynamical instability of giants would change the stable region of test particles in Section~\ref{results_freshtps}.
Our results are compared with the observations in Section~\ref{results_obs}.  
\subsection{Fates of Test Particles Depending on Minimum Orbital Distance of
Jupiters}\label{results_general}
We evolve each of 40 systems up to 10\,Myr.   
When systems have two Jupiters and more than one survived test particles at the end of the simulations, we run them up to 100\,Myr.   
However, the overall results do not change significantly. 

\begin{figure}
\plotone{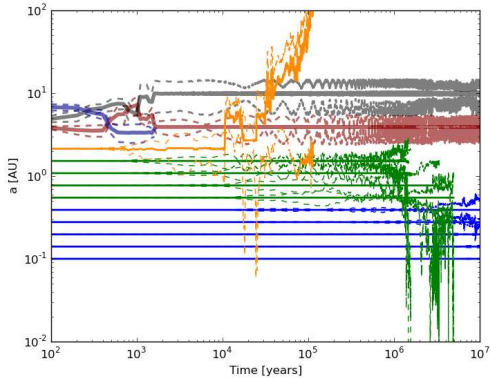}
\caption{Orbital evolution of three Jupiter-mass planets (thick curves) and
eleven test particles (thin curves).   
Note that one of the test particles hits the star before 100\,yr and thus does
not appear in this plot. 
Dashed curves indicate peri- and apo-centers of each particle.
The orange, green, and blue colors of test particles indicate their fates of
ejection, merger with the central star, and survival for 10\,Myr, respectively. 
\label{run36_all}}
\end{figure}
An example of the orbital evolution of our model is shown in Figure~\ref{run36_all}. 
Three thick curves and eleven thin curves correspond to semi-major axes of 
Jupiter-mass planets and test particles, respectively.  Dashed curves represent each particle's peri- and apo-center.  
This example shows a relatively ``quiet'' dynamical evolution of giant planets, 
where their orbital radii and eccentricities do not change much -- the outermost Jupiter-mass planet crosses orbits with inner two giants, 
and collides with the innermost Jupiter-mass planet around $2000\,$yr. 
On the other hand, test particles appear to take various dynamical outcomes, largely depending on their distances from Jupiters.   
The outermost test particle hits the central star immediately after the start of the simulation ($\sim36\,$yr), and thus is not shown in this plot. 
For the rest of the test particles, the outer test particle (orange curve) experiences orbital crossings with giants and is ejected out of the system (referred to as ``{\it ejected}'' particles), while the inner test particles (blue curves) stay dynamically stable for 10Myr (``{\it survived}'' particles). Over the intermediate radii, test particles which do not have orbital crossings with giant planets merge with the central star (green curves; ``{\it merged}'' particles).  
Although not shown in this example, when test particles collide with giant planets, we call them ``{\it collided}'' particles and plot them in red.

\begin{figure}
\plotone{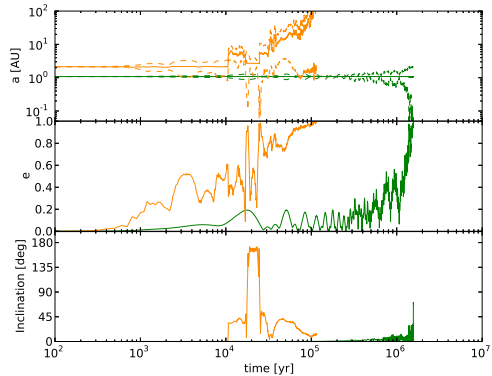}
\caption{Orbital evolution of a merged and an ejected test particle in
Figure~\ref{run36_all}. The 8th and 10th test particles from the star are shown
with green and orange curves, respectively.  Top, middle, and bottom panels show
evolution of semimajor axes, eccentricities, and inclinations, respectively. 
The dashed curves in the top panel are peri- and apo-center of the corresponding orbits.   
For the ejected test particle, both eccentricity and semimajor axis change
within a short period of time due to strong gravitational interactions with
Jupiter-mass planets. 
On the other hand, for the merged test particle, its semimajor axis stays
constant while the eccentricity increases due to secular interactions, which leads
to a small pericenter distance and an eventual merger with the central star. 
\label{run36_tp7_tp9}}
\end{figure}
When the test particles have orbital crossings with giant planets, 
they tend to experience close encounters and strong interactions, which could dramatically change the orbits of test particles. 
On the other hand, even when the test particles do not have orbital crossings with giants, their orbits could still be perturbed by secular interactions.
In Figure~\ref{run36_tp7_tp9}, we highlight these differences between ejected and merged test particles.
Since the ejected test particle (orange curves) has orbital crossings 
with giants, the semi-major axis changes rapidly while the eccentricity and inclination increase.
On the other hand, for a merged test particle shown here (green curves), secular interactions dominate its evolution, 
and thus semi-major axis stays constant while the eccentricity increases. 
Figures~\ref{run36_all} and \ref{run36_tp7_tp9} indicate that, since secular interactions do not change a semi-major axis, 
the increase of an eccentricity leads to a smaller and smaller pericenter of a test particle and an eventual merger of the particle with the central star, 
rather than the ejection out of the system.

\begin{figure}
\plotone{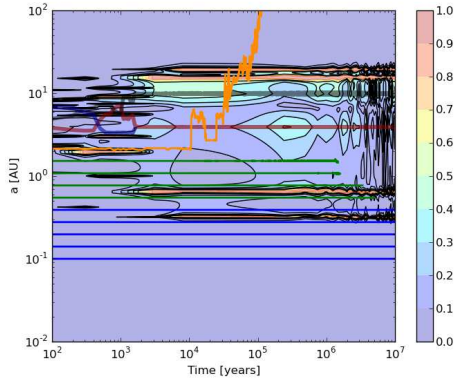}
\caption{Same as Figure~\ref{run36_all}, but without peri- and apo-centers, and
with forced eccentricity contours which are estimated by the lowest-order
secular theory 
with the orbital parameters of Jupiter-mass planets at each time.  Orbits of
merged test particles (green lines) overlap with the high forced-eccentricity regions. 
\label{run36_all_forced}}
\end{figure}
To confirm the effects of secular interactions on orbital evolution of test particles, we compare the expectations from the secular theory with the simulated results.  
We adopt the lowest-order secular theory, and thus assume small eccentricities and inclinations for the planets. 
This is a rather crude assumption, because eccentricities and inclinations of giant planets can become very large as a result of the dynamical instability.  
Furthermore, since test particles are distributed over $0.1-3\,$AU in our model, their eccentricities must increase above 0.95 to hit the Sun-like central star.
Nevertheless, as shown below, we find that the lowest-order secular theory predicts the fates of test particles fairly well.

For the lowest-order secular theory, the evolutions of eccentricity and inclination are decoupled.  
Here, we focus on the eccentricity evolution, because that determines whether a test particle merges with the star or not. 
The secular solutions for eccentricity evolution consist of free and forced components.  
%
\begin{eqnarray}
e\sin\varpi &=& e_{\rm free}\sin\left(At+\beta\right) + e_{\rm forced}\sin\varpi_{\rm forced} \\
e\cos\varpi &=& e_{\rm free}\cos\left(At+\beta\right) + e_{\rm forced}\cos\varpi_{\rm forced}
\end{eqnarray}
Here, $e$ is the eccentricity, $\varpi$ is the longitude of pericenter, and $\beta$ is the phase.
The free component is characterized by a free frequency $A$ of the test particle's apsidal precession due to gravitational interactions with giant planets, while the forced component is characterized by N eigen frequencies $g_i$ of an N-giant-planet system:  
\begin{equation}
e_{\rm forced}\sin\varpi_{\rm forced} =  - \displaystyle\sum\limits_{i=1}^N \frac{\nu_i}{A-g_i}\sin\left(g_it+\beta_i\right) \label{eforced} \, . 
\end{equation}
Here, $\nu_i=\Sigma_{j=1}^{N}A_je_{ji}$, where $e_{ji}$ is a component of an unscaled eigen vector which corresponds to $g_i$, and $A_j$ depends on masses of a giant planet and the central star as well as the orbital radii of the giant planet and a test particle \citep[for details, see][]{Murray99}. 
A similar equation holds for the cosine term.
These equations indicate that the secular interactions become the strongest near resonance locations, where 
a free frequency of a test particle becomes comparable to one of the eigen frequencies of these giants $A\sim g_i$.   
Since the efficient angular momentum exchanges lead to high forced eccentricities near secular resonances, we expect that test particles in these regions are prone to mergers.

In Figure~\ref{run36_all_forced}, we compare the evolution of forced eccentricity contours with the orbital evolution shown in Figure~\ref{run36_all}.  
The orbits of merged test particles coincide with high forced eccentricity regions, while those of survived test particles are in small forced eccentricity regions. 
We have performed similar comparisons for all of our simulations, and found that the forced eccentricities of test particles are good indicators of removed and survived planets.

\begin{figure}
\plotone{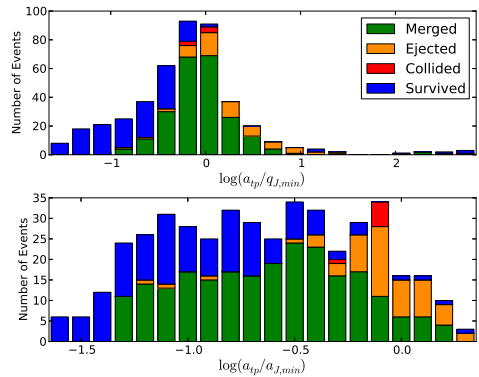}
\caption{Fates of test particles depending on the initial semimajor axes of test
particles that are normalized by the minimum pericenter distances (top panel) 
and the minimum semimajor axes of Jupiter-mass planets (bottom panel). 
The number of the ejected test particles increase toward and beyond $a_{tp,in}/a_{J,min} = 1$, while most merged particles are within this limit.
\label{hist_peri_fate_qmin_amin}}
\end{figure}
%
If the secular interaction is a dominant cause of merged particles while the close encounters is a dominant cause of ejected particles, we would expect that the fates of test particles are divided clearly depending on whether test particles cross orbits with giant planets or not.
We test this hypothesis in Figure~\ref{hist_peri_fate_qmin_amin}, where the test particles' initial semimajor axes $a_{tp,in}$ 
are compared with the minimum pericenter $q_{J,min}$ (top panel) and semimajor axes $a_{J,min}$ (bottom panel) of giant planets.  
Note that, the ``minimum'' pericenter (or semimajor axis) of a Jupiter can be different 
for each test particle in the same simulation, because this is chosen during each test particle's survival time 
(i.e., if the test particle survived until the end of the simulation, 
the minimum semi-major axis would be that throughout the simulation).
Surprisingly, we find that about half of the merged test particles have $a_{tp,in}/q_{J,min}\geq 1$, 
and thus are potential candidates for orbital crossings (see top panel). 
By checking the orbital evolution of planets for each simulation, we find that some merged test particles have clear orbital crossings with giant planets, just like ejected test particles.  
However, most merged test particles have nearly constant semimajor axes until the merger times, and thus are expected to have negligible close encounters with giant planets.

\begin{figure}
\plotone{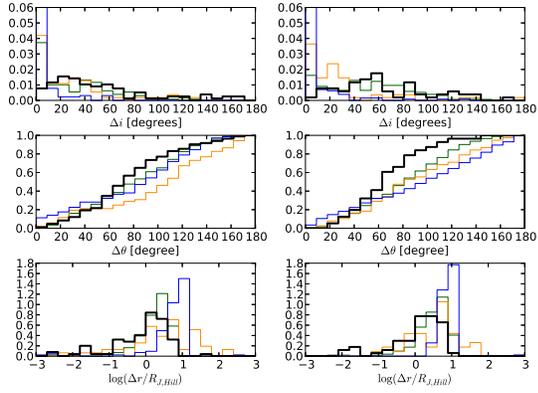}
\caption{
Comparisons of the mutual inclinations (top panels), angles between pericenters (middle panels), and minimum orbital distances (bottom panels). 
Left panels show the results when the pericenter of the innermost giant planet takes the smallest value $q_{J,min}$, while right panels show the corresponding results at the end of the test particles' lifetimes (or at the end of the simulations for survived ones).  
The color codes are the same as Figure~\ref{hist_peri_fate_qmin_amin}, except that
the merged test particles with $a_{tp,in}/q_{J,min}\geq1$ are plotted in black instead of dark green.
The middle panels show the cumulative histograms of the angles between pericenters $\Delta\theta$ for each case corresponding to the top panels.  
The bottom panels show the minimum distance between orbits of test particles and the innermost giant planets. This distance peaks near 10 Hill radii of the innermost giant planets for the survived test particles, and near a few Hill radii for both merged and crossed-then-merged particles.  
\label{periang_aratio}}
\end{figure}
%
Figure~\ref{periang_aratio} shows that the ``crossed-then-merged'' (merged with $a_{tp,in}/q_{J,min}\geq1$) test particles actually avoid close encounters 
with giants due to large mutual inclinations between orbits, and/or a tendency toward an orbital alignment.
From the top left panel, we find that, when $q_{J,\min}$ is reached, the mutual inclination between orbits peaks near $0^{\circ}$ for all cases except for the crossed-then-merged ones.  The clear tendency toward large mutual inclinations for the crossed-then-merged cases become stronger at the end of the simulations (top right panel).

For the middle panels, we introduce an angle between pericenters of two orbits $\Delta\theta$.   We use this angle as an indicator of the orbital alignment instead of the angles between argument of pericenters $\Delta\varpi$, because $\Delta\varpi\sim0^{\circ}$ does not necessarily mean an alignment of pericenters depending on the mutual inclination.
In both left and right panels, we find that survived test particles tend to be more aligned ($\Delta\theta\sim0^{\circ}$) than the others. The crossed-then-merged cases do not show such a strong tendency toward the alignment, but they avoid the misalignment ($\Delta\theta\sim180^{\circ}$) which would place the apocenter of the inner test particle close to the pericenter of the outer giant planet, and therefore could result in strong interactions between them.  This tendency {\it away} from the misalignment becomes stronger toward the end of the simulations (middle right panel), where the $\Delta\theta$ peaks near $70^{\circ}$.
Therefore, these two sets of panels indicate that crossed-then-merged particles could have  avoided close encounters with giants due to large mutual inclinations and/or relatively well-aligned orbits. 

To check this point further, we compare the minimum distances between the orbits of a test particle and the innermost giant planet in the system (bottom panels). As expected, this distance is the largest for the survived test particles, and smaller for merged particles.  
For the ejected test particles, the distribution is broader, which indicates that many of these test particles started crossing orbits with giant planets before $q_{J,min}$ was reached.
For the survived test particles, the minimum orbital distance peaks near 10 Hill radii of the giant planet, while for the merged and crossed-then-merged particles, this peaks near a few Hill radii.
Although some crossed-then-merged test particles have very small minimum orbital distances ($\Delta r < 1\,$ Hill radius), the overall similarity of the distributions of merged and crossed-then-merged cases indicates that many crossed-then-merged particles have avoided close encounters with giant planets. 

From these results, we conclude that $a_{tp,in}/a_{J,min}$ is the better indication of close encounters compared to $a_{tp,in}/q_{J,min}$, 
since the former shows the transition from merger-dominated to ejection-dominated cases near $a_{tp,in} = a_{J,min}$, while the latter does not. 
The bottom panel of Figure~\ref{hist_peri_fate_qmin_amin} shows that the fates of test particles depend on distances from giants -- from the innermost region to outward, survived, merged, and ejected test particles are distributed.
Our results show that secular perturbations from Jupiters can remove test particles down to at least $\sim30$ times smaller than their minimum semimajor axis, or $\sim10$ times smaller than their minimum pericenter distance.
This corresponds to the case where a test particle at $0.1\,$AU is removed by giant planets beyond $3\,$AU.
Note that this lower limit is not definite, because we do not have test particles all the way down to the central star.  
Another notable point is a large fraction of the merged particles compared to the ejected ones.  
Even in the ``orbit crossing'' cases with $q_{J,min} < a_{tp,in} \; (< a_{J,min})$, merger is dominated over ejection. 
As a reference, the percentages of survived, merged, ejected, and collided test particles are $32.5\,\%$, $52.0\,\%$, $13.9\,\%$, and $1.6\,\%$, respectively.
This is a very different outcome from \cite{Veras05}, where $>95\,\%$ of test particles are ejected.  The difference could be caused by the fact that \cite{Veras05} focused on the stability in HZ, while we are simulating test particles over a wider range of orbital radii.
Recently, a correlation between the existence of debris disks and that of terrestrial planets were studied by \cite{Raymond11,Raymond12}. Their orbital calculations showed that the dynamical instabilities among giant planets drove most inner planets toward the star. This is consistent with our result, although they stopped calculations when the planets entered the region within $0.2\,$AU.
%
%
%
\subsection{Estimating Dynamically Unstable Region of Test Particles}\label{results_stability}
In the last subsection, we have found that the secular interaction is a dominant cause of the mergers of test particles with the central stars.   
In this subsection, we test this hypothesis and propose a convenient formula which allows us to estimate the dynamically unstable region of the test particles from the orbital evolution of giant planets alone.

As we have seen in the last subsection, test particles need to achieve $e\sim1$ to hit the central stars.  On the other hand, in the secular theory, the eccentricity is determined by the vector sum of the free and forced eccentricities.   
In the plane of $(e\cos\varpi,\,e\sin\varpi)$, we can think that the free eccentricity vector completes a circular motion around the forced eccentricity vector $(e_{\rm forced}\cos\varpi_{\rm forced},\,e_{\rm forced}\sin\varpi_{\rm forced})$ at the free frequency $A$ while the forced vector itself is also moving \citep[e.g.,][]{Murray99}.
Thus, we can expect that when the forced eccentricity of a test particle stays above 0.5 for the time longer than a secular timescale, the eccentricity may reach $\sim1$.

\begin{figure}
\plotone{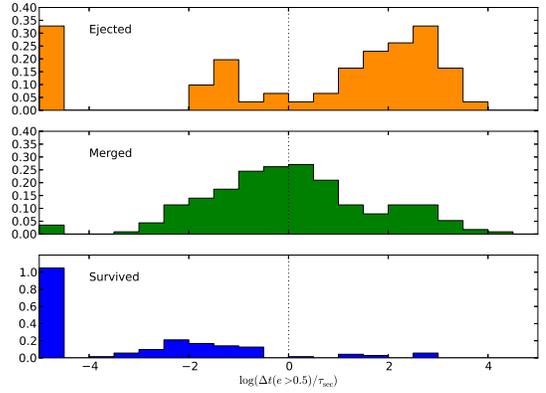}
\caption{Comparison of the ratios of the total durations of the forced eccentricities of test particles being above 0.5 and the secular timescales.  The forced eccentricities are estimated from the orbital parameters of Jupiters.  This time ratio gives a good prediction of whether test particles being dynamically unstable or not. 
Most ejected and merged test particles exist beyond the ratio 1, while most survived test particles are below this limit.
A peak at $10^{-5}$ for survived test particles represents all the particles with $\Delta t(e>0.5)/\tau_{\rm sec} < 10^{-4.5}$
that do not achieve the estimated forced eccentricities above 0.5 throughout the simulations.  
\label{tecc_tsec}}
\end{figure}

In Figure~\ref{tecc_tsec}, we test this hypothesis by calculating the forced eccentricity of a test particle from the orbital parameters of giant planets and comparing the total duration of $e_{\rm forced}>0.5$ with a secular time:  
\begin{equation}
\tau_{\rm sec} = \min\left(2\pi/A,\,2\pi/g_i\right) \ .
\end{equation}
Here, $g_i$ is an eigenfrequency of giants, and we choose the frequency that corresponds to the largest coefficient $\nu_i/(A-g_i)$ in Eq.~\ref{eforced}.

For the survived planets, we expect that the eccentricities rarely reach 0.5, and thus $\Delta t(e>0.5)/\tau_{\rm sec}\ll1$, which agrees well with the bottom panel of Figure~\ref{tecc_tsec}.
On the other hand, we do not expect any particular correlation for the ejected particles, because we calculate $\Delta t(e>0.5)$ by assuming the initial semimajor axes of test particles, and the semimajor axes of ejected test particles often change significantly over time due to orbital crossings and close encounters with giants (see, e.g., Figure~\ref{run36_all}).
However, the top panel of the figure indicates that $\Delta t(e>0.5)/\tau_{\rm sec}$ peaks near $10^2-10^3$ for the ejected test particles.
This suggests that it takes a few hundred secular times before these test particles are finally ejected.  Note that, since most ejected test particles are in the outer region of the system, their secular timescales tend to be short ($\sim10^3-10^4\,$yr).
For the merged test particles, we expect that the distribution peaks sharply near $\Delta t(e>0.5)/\tau_{\rm sec}\sim1$.
The middle panel of the figure shows that the distribution indeed peaks near 1, but also spreads broadly, from $\sim10^{-3}$ to $\sim10^4$.
We find that most merged test particles around the bump near $10^2-10^3$ act similar to the ejected particles, and have varying semimajor axes or even orbital crossings with giants.
On the other hand, for $\Delta t(e>0.5)/\tau_{\rm sec}<10^{-1}$, test particles tend to merge with the central star as one of the giants obtains a very small pericenter distance, either on its way toward an ejection out of the system, or a merger with the central star. Since these giant planets tend to have very high eccentricities, it is likely that we are overestimating the secular timescale for these cases.
From these results, we find the following trends, which hold for Set\,1 and 2 as well.  
%
%
\begin{eqnarray}
\Delta t(e>0.5)/\tau_{\rm sec} &\gtrsim& 10 \, {\rm :\, ejected} \\
							   &\sim& 10^{-2}-10 \, {\rm :\, merged} \\
                               &\lesssim& 10^{-2} \, {\rm :\, survived}
\end{eqnarray}
%
%
\subsection{Fates of Test Particles Depending on Dynamical Outcomes of
Jupiters}\label{results_Joutcomes}
In the previous sections, we have discussed the trends seen from all the simulations.  
We have shown that the fates of test particles are determined by secular interactions as well as close encounters with giant planets.
Therefore, it is natural to expect that the fates of test particles depend on dynamical outcomes of giant planets as well. We explore this possibility here.

\begin{figure}
\plotone{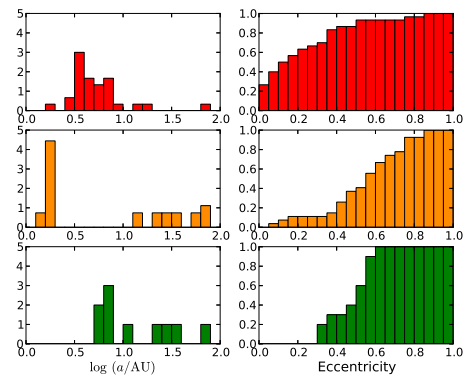}
\caption{Left: normalized semi-major axis distributions for the survived Jupiter-like planets.  From top to bottom, Jupiter-Jupiter collision, Jupiter ejection, and Jupiter-star merger cases. Right: normalized, cumulative distributions of corresponding eccentricities. 
\label{hist_semia_ecc}}
\end{figure}
%
By the end of the simulations, all forty systems have lost at least one Jupiter-mass planet, and thirteen systems have lost two Jupiters.  
When we categorize the systems by the first-lost giant planet, $17/40$, $15/40$, and $8/40$ systems have Jupiter-Jupiter collisions, Jupiter ejections, 
and Jupiter-star mergers, respectively.   
As shown in Figure~\ref{hist_semia_ecc}, for Jupiter-Jupiter collision cases, semi-major axes of survived Jupiter-like planets tend to be similar to the initial values, and more than $90\,\%$ of these planets have eccentricities less than 0.5. 
On the other hand, for Jupiter ejection and Jupiter-star merger cases, final semi-major axes of giant planets are more widely distributed, and less than $40\,\%$ of planets have eccentricities below 0.5.  
This plot indicates that, as expected, Jupiter-Jupiter collision cases are relatively dynamically quiet, while ejection and merger cases are more violent. 

\begin{figure}
\plotone{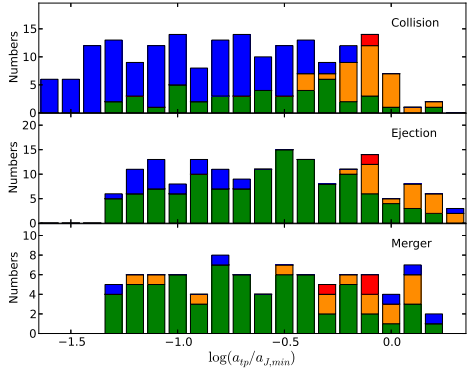}
\caption{Dependence of test particles' fates on dynamical outcomes of giant
planets.  Each panel shows the fates of test particles as a function of the initial
semimajor axes 
of test particles normalized by the minimum semimajor axes of Jupiter-mass
planets.  Top, middle, and bottom panels correspond to the systems with 
collisions of two giant planets, ejections of giant planets, and mergers of
giant planets with the central stars.  It is apparent that most survived test
particles reside in 
Jupiter-Jupiter collision cases, while almost all test particles are gone for
Jupiter-star merger cases.
\label{hist_peri_fate_CEMonly_amin}}
\end{figure}
In Figure~\ref{hist_peri_fate_CEMonly_amin}, we show histograms similar to the bottom panel of Figure~\ref{hist_peri_fate_qmin_amin}, depending on Jupiters' dynamical outcomes.  
As shown in Figure~\ref{hist_semia_ecc}, when giant planets collide with each other, neither semi-major axes nor eccentricities of these planets change largely.
Such evolution of giant planets should have the least significant effects on orbital stability of test particles.  
Indeed, the top panel shows that about three quarters of the survived test particles belong to the Jupiter-Jupiter collision cases.  

On the other hand, when a Jupiter merges with the central star (bottom panel), the planet sweeps through the area of test particles and removes most of them.
There are several survived test particles as well.  However, as we can guess from their random distribution in $a_{tp,in}/a_{J,\min}$, all of them have high eccentricities.    
Note that, in this Jupiter-star merger case, most test particles also merge with the central stars rather than being ejected out of the systems. 
Thus, although giant planets approach test particles, the eccentricities of these particles are excited not by close encounters, but by secular interactions.
 
The middle panel shows the results of the Jupiter-ejection cases.  Here, the pericenters of ejected Jupiters do not necessarily come close to the regions of test particles.  Therefore, some test particles can survive.  However, the majority of test particles either merge with the central stars or are ejected. 

For Jupiter-Jupiter collision, Jupiter ejection, and Jupiter-star merger cases, the survival percentage of test particles are $61\,\%$, $14.5\,\%$, and $5.7\,\%$, respectively.
Thus, our results suggest that, unless the evolution of giant planets is either stable or quiet (e.g., Jupiter-Jupiter collision cases), most test particles interior to 
Jupiters' orbits would be removed from the systems, even if their orbital radii are several times smaller than
the pericenter distance of the innermost survived Jupiters.   
Since systems which underwent dynamically violent evolutions tend to have giant planets on eccentric orbits, our results suggest that systems with eccentric Jupiters at a few AU may have lower chances of hosting terrestrial planets down to ~0.1AU.
We discuss this issue further in Section~\ref{summary}.
%
%
%
\subsection{Effects of Dynamical Instability on Stable Regions}\label{results_freshtps}
Previous studies estimated the stability of hypothetical terrestrial planets in the HZs of the observed systems \citep[e.g.,][]{Gehman96,Jones01,Noble02,Menou03,Asghari04}. 
However, since these studies did not take account of the past evolution of the observed systems, 
they are likely to have overestimated the dynamically stable region of terrestrial planets.   
In the cases of Jupiter ejections and and Jupiter-star mergers,
Jupiters with the highest eccentricities must have disappeared from the systems and
these planets should have strongly destabilized the orbits of terrestrial planets through
secular perturbations that were highly enhanced by the largest eccentricities.
In this subsection, we show how the estimated stable region would change by ignoring the past orbital evolution.
To study this, we add the same 11 test particles we used as initial conditions to the final state of giant planets for each simulation, 
and run the simulations for $10\,$Myr.   
As expected, we find that many more test particles survive until the end of the simulations.

\begin{figure}
\plotone{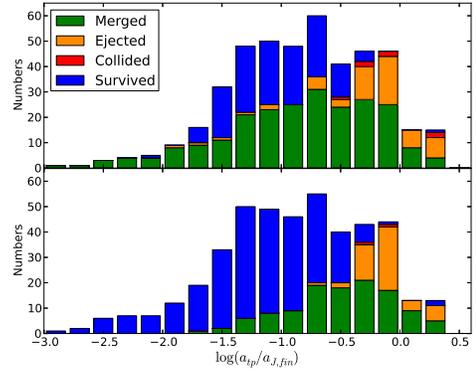}
\caption{The effects of dynamical instability of giant planets on the orbital stability of test particles.  
The top panel is similar to the bottom panel of Figure~\ref{hist_peri_fate_qmin_amin}, but the initial semimajor axis of a test particle is normalized by the final semimajor axis of the innermost giant planet.
The bottom panel shows the corresponding results for simulations of 11 test particles added to the final states of giant planets. 
Many more test particles survive in the latter case compared to the former,
which indicates that dynamical instability among giant planets indeed affects
the stable region of the test particles. 
\label{hist_peri_fate_afin_afterwards}}
\end{figure}
Figure~\ref{hist_peri_fate_afin_afterwards} compares the outcomes of two sets of simulations. 
The top panel is similar to the bottom panel of Figure~\ref{hist_peri_fate_qmin_amin}, but shows the fates of test particles in terms of the initial semi-major axes of test particles normalized by the innermost Jupiter's semimajor axes at the end of the simulations $a_{tp,in}/a_{J,fin}$, instead of $a_{tp,in}/a_{J,\min}$.   
The corresponding figure for the new set of simulations is shown in the bottom panel.

In the top panel, we see the tail of ejected test particles with small $a_{tp,in}/a_{J,fin}$.  
This tail is a result of different normalization factors (i.e., $a_{J,fin}$ instead of $a_{J,\min}$), and does not mean that there are many ejected particles interior to the stable test particles. 
Most of the ejected test particles in this tail belong to the Jupiter-star merger cases, where the survived giant planets often have large final semimajor axes (see Figure~\ref{hist_semia_ecc}).
  
While the overall shapes of these two distributions are similar, there is a significant increase in the fraction of stable test particles in the bottom panel, 
which confirms our expectation that past dynamical instabilities among giants strongly influence the orbital stability of test particles.
In other words, the stable regions of hypothetical terrestrial planets that are determined 
from the {\it current} distributions of giant planets would be largely overestimated.

For the new simulations, the fractions of the survived test particles have increased for all kinds of giant planet evolutions, compared to the original simulations.  
As expected, the increase is most significant for Jupiter-star merger cases, and very modest for Jupiter-Jupiter collision cases. 
The large increase in the number of survived test particles for Jupiter-star merger and Jupiter ejection cases indicates that, even when the stability analysis allows the existence of stable hypothetical terrestrial planets, few such planets would exist interior to moderately to highly eccentric giant planets.  
However, we should note that our simulations do not take account of tidal circularization effects, which may have saved some of these terrestrial planets.  We briefly discuss this possibility in Section~\ref{summary}.
%
%
%
\subsection{Comparisons with Observations}\label{results_obs}
\begin{figure}
\plotone{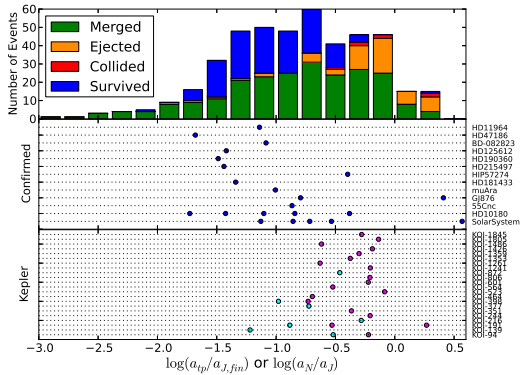}
\caption{Comparison of our results shown in the top panel of
Figure~\ref{hist_peri_fate_afin_afterwards} with the observed multiple-planet systems. 
For the middle panel, we select 
all systems with both Neptune- and Jupiter-like planets, where these planets are
defined to be less and more massive than $0.1M_J$.  
The ratios of these planets' semimajor axes to the innermost Jupiter-like planets' semimajor axes are plotted. 
The observed systems fall in the stable test-particle area that is estimated from
our simulations.
For the bottom panel, Kepler's multiple-planet-candidate systems are shown. All
systems with planets smaller and larger than $6R_E$ are selected. 
The cyan circles show the smaller planets that have the radii comparable to Earth of $\leq 2R_E$, while the pink circles represent the rest.   
Kepler planets tend to be much closely separated from each other, compared to
the expectations from our numerical simulations. 
\label{hist_peri_fate_exoplanets}}
\end{figure}

In this subsection, we compare our numerical results with the distribution of the observed exoplanetary systems.
Out of over 600 confirmed planetary systems, more than 100 are multiple-planet systems 
(\url{http://exoplanet.eu/}).  We choose systems with both Neptune- and Jupiter-mass planets, where we define these planets to have less and more masses than $0.1M_J$, respectively. 
There are 13 systems like that including our solar system, and the semimajor axis ratios 
for Neptune- and Jupiter-mass planets are plotted in the middle panel of Figure~\ref{hist_peri_fate_exoplanets}.
The eccentricities of giant planets in these systems are lower than the other giant planets, and more than $90\,\%$ of them have eccentricities less than 0.4. 
The preference for low eccentricities of giant planets in mixed-size multiple-planet systems agrees with the expectation from our simulations (see Section~\ref{results_Joutcomes}).

In Figure~\ref{hist_peri_fate_exoplanets}, the circles on the same dotted line belong to the same system. 
We find that the confirmed systems with both Neptune- and Jupiter-like planets are distributed within the range of survived test particles from our simulations (top panel).
There are two circles exterior to this region, and beyond $a_N/a_J=1$.
However, since our simulations do not have test particles outside the Jupiters' orbits initially, we do not intend to estimate the stable region beyond $a_N/a_J=1$.
%

We perform a similar comparison for Kepler's planet candidates as well, by choosing systems with Neptune- and Jupiter-size planets (smaller and
larger than $6R_E$, respectively).   
The limit is adopted so that it is consistent with the classification of Kepler's planetary candidates \citep[e.g.,][]{Batalha12ap}. 
Out of about 360 multiple-planet-candidate systems, 22 have both of these planets.
As shown in the bottom panel of Figure~\ref{hist_peri_fate_exoplanets}, 
Kepler's planet candidates turn out to be more closely separated from each other than our simulations predict. 
However, the agreement is better for systems with smallest planets of $\leq2R_E$ (cyan circles).  
There are a few potential reasons why the distributions of Kepler's planet candidates are less consistent with our simulations.
First, it is possible that most multiple-planet-candidate systems did not experience any dynamical instability. 
If this is the case, we would expect that these planetary candidates have small eccentricities.  The future observations would help clarifying this possibility. 
Secondly, the agreement may improve if we take masses of small planets into account. 
We discuss this point further in the next section.
Thirdly, our simulations are done with equal-mass giant planets, and thus are expected to place the lower limit on the dynamically stable region of test particles. 
It is possible that many Kepler planet candidates are actually in the ``stable'' region. 
%
%
%
%
%
\section{Discussion and Conclusions}\label{summary}
The recent observations discovered many multiple-planet systems, including the ones with multiple sub-Neptune-size planets on short-period orbits.
Furthermore, there are many giant-planet systems which allow the presence of terrestrial planets in HZs.
However, it is still unclear whether there are many planetary systems 
in which giant and terrestrial planets coexist like the solar system.
In fact, Kepler and HARPS surveys 
suggest that close-in sub-Neptune-size planets
are generally not accompanied by a gas giant planet in the same system -- $\sim5\,\%-10\,\%$ of multiple-planet systems have such mixed-size planets. 
 
One of the potential causes that prevents the coexistence of giant and terrestrial planets would be early evolution of giant-planet systems.
The migration and dynamical instability of giant planets are two potentially important effects on formation and survival of terrestrial planets.   
\citet{Veras05} pointed out that when dynamical instability of giant planets occurs,
terrestrial planets in the HZ undergo close encounters with the giants and are ejected
out of the system.

In this paper, we have studied how dynamical instability among 
gas giant planets affects the orbital stability of sub-Neptune-like planets interior to their orbits, particularly that of close-in terrestrial  planets.  
We have performed N-body simulations of systems with three Jupiter-mass planets
and test particles corresponding to terrestrial planets over $0.1-3\,$AU.
The important results we have found are as follows: 
\begin{itemize}
\item{
Dynamical instability often leads to either an ejection of a giant planet or a merger of such a planet with the central star.  
In these cases, test particles are removed over a broad range of semimajor axis $a_{tp,in}$ down to $\sim10$ times smaller than the minimum pericenter distance of the giants $q_{J,\min}$ (see Figure~\ref{hist_peri_fate_qmin_amin}).
This implies that stable regions evaluated by the current distribution of giants is significantly overestimated.
}

\item{
A major cause of removal of test particles is secular perturbations from giant planets.
Eccentricity excitement of giants during their orbit crossing highly enhances the
effect of secular perturbations. 
Since secular perturbations continuously increase eccentricities of the test particles
without changing their semimajor axes, most of the removed particles
merge with the host star rather than being ejected out of the system.
}

\item{
Even for orbit crossing cases of $a_{tp,in}/q_{J,\min} \geq 1$, there are many mergers (top panel of Figure~\ref{hist_peri_fate_qmin_amin}).
Most of these test particles do not have actual close encounters with giant planets, due to large mutual inclinations between the orbits, 
and/or tendency {\it away} from the pericenter misalignment (see Figure~\ref{periang_aratio}). 
Due to these effects, the merged and ejected particles are divided around $a_{tp,in}/a_{J,\min} = 1$, rather than $a_{tp,in}/q_{J,\min} = 1$. 
}

\item{
Since most test particles are removed as a result of secular interactions but not of close scattering, we can semi-analytically estimate the dynamically unstable region of test particles (see Section~\ref{results_stability}). 
We have found that the ratio of the total duration of the forced eccentricity $e_{\rm forced}>0.5$ and the secular time $\Delta t(e>0.5)/\tau_{\rm sec}$ is 
a good measure for this region (see Figure~\ref{tecc_tsec}).
This formula allows us to estimate the unstable region of test particles just from the simulations of giant planets.
}
\end{itemize}

When the orbital instability of gas giants does not occur
or only Jupiter-Jupiter collisions occur, 
secular perturbations do not remove terrestrial planets over a wide range of orbital radii because the eccentricities of giants stay low.
On the other hand, in the systems with giants of relatively high eccentricities, the secular perturbations are highly enhanced and the enhanced regions migrate back and forth due to orbit crossings so that most terrestrial planets interior to the orbits of giants are removed except for ones very close to the host stars.
Our work further indicates that the eccentricities of survived particles
are not high, since secular perturbations sensitively depend on the ratio of semimajor axes of a test particle and giant planets. 
Our theoretical prediction of the orbital stable region is consistent with the observational data (see Section~\ref{results_obs}).

Note that there are some caveats and limitations in our analysis.
For example, the lowest-order secular theory used for the semi-analytical formula, in principle, cannot be applied to the high eccentricity cases.
However, we have found that it still provides a fairly good agreement with the numerical simulations. 

Also, our initial conditions may appear to be rather specific to draw a general conclusion. To clarify this point, we perform two more sets of simulations, as described in Section~\ref{method} --- Set\,1: one giant planet has a lower mass ($0.5\,M_J$), and Set\,2: larger initial separations, eccentricities, and inclinations for giant planets.  We have performed 42 simulations for the former set, and 40 for the latter.  In both sets, the overall results are similar to the default set --- test particles are removed over a wide range of orbits, and at least down to $\sim10$ times smaller than the minimum pericenter distance of giant planets, or in other words, $\sim30$ times smaller than the minimum semimajor axis of these planets.  The trends of $\Delta t(e>0.5)/\tau_{\rm sec}$ we discussed in Section~\ref{results_stability} are also consistent with both of these sets.

There are some differences between these sets and the default set.  For Set\,1, 17/42, 22/42, and 3/42 systems have Jupiter-Jupiter collisions, Jupiter ejections, and Jupiter-star mergers, where the most common outcome of giant planet evolution is the ejection of the low-mass giant planet\footnote{Although one of the giant planets is now half a Jupiter mass, we conventionally use "Jupiter-Jupiter" collisions etc..}.  The percentages of survived, merged, ejected, and collided test particles are $40.0\,\%$, $45.9\,\%$, $12.8\,\%$, and $1.3\%$.
Thus, the total survival percentage of test particles $40\%$ is higher than our default case of $32.5\%$, despite that there are more ejections and mergers of giants, and thus many more "dynamically-unstable" systems in Set\,1.  Furthermore, we find that the survival percentage is $\sim10\%$ higher in Jupiter-Jupiter collision cases, and $\sim3\%$ higher in Jupiter ejection cases, compared to the default set. This suggests that the forced eccentricities do not reach above 0.5 often in Set\,1. By comparing the evolution of forced eccentricities similar to Figure~\ref{run36_all_forced}, we find that the variations of forced eccentricities are smaller in Jupiter-Jupiter collision cases. Thus, Set\,1 demonstrates that equal-mass planet systems are likely to give the lower limit on the survival rate of test particles, agreeing with the previous studies like \cite{Veras06} and \cite{Raymond12}. However, we do not find any significant difference in the distributions of fates of test particles such as Figures~\ref{hist_peri_fate_qmin_amin} and \ref{hist_peri_fate_CEMonly_amin}.  We may find a larger stability region for the greater mass difference. 

For Set\,2, 3/40, 27/40, and 10/40 systems have Jupiter-Jupiter collisions, Jupiter ejections, and Jupiter-star mergers.  Much lower occurrence of Jupiter-Jupiter collision cases is due to the higher eccentricities and inclinations assumed for the initial distributions of Jupiters. The percentages of survived, merged, and ejected test particles are $16.1\,\%$, $75.7\,\%$, and $8.2\%$. There are no test particle-planet collisions in this set.  The low survival rate is due to the small number of Jupiter-Jupiter collision cases.  Note, however, that the distributions of fates of test particles are similar to the default set.  Thus, Set\,2 demonstrates that the range of a stable region, and thus the final outcome of dynamical evolution of plants, is not largely affected once the dynamical instability sets in.  We also note that the survival percentage of test particles shown in this paper should be treated as a reference value, and not a definite one.  As stated in Section~\ref{intro}, to obtain a more reliable survival rate, we need to start with the giant planet distribution that reproduces the observed planet distribution.   

Another assumption in our study is that we treat sub-Neptune-size planets as test particles.  The self-gravity among these planets may inhibit instability, because gravitational interactions among the massive particles that are perturbed by giant planets tend to align neighboring particles with one another and make their eccentricity evolution synchronized \citep[e.g.,][]{Ito01,Levison03}. 
On the other hand, such an effect can also lead to mergers of all the planets with the central star all at once. 

\begin{figure}
\plotone{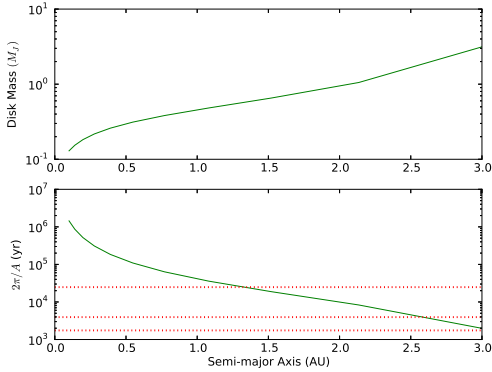}
\caption{Top: estimate of a disk mass below which the eccentricity excitation cannot be ignored, 
because a secular timescale becomes shorter than an eccentricity damping timescale of a disk. 
Bottom: comparison of secular timescales. The green curve corresponds to a free frequency $A$ of a test particle, while the red dotted lines are from the three eigen values $g_i$ of Jupiter planets.  
The eccentricity increase is most significant near resonance locations where these lines intersect. 
\label{tsec_tedamp_sigma}}
\end{figure}
Furthermore, we have implicitly assumed that dynamical instabilities among giant planets occur {\it after} the gas disk dissipates, 
although dynamical instabilities could take place once the disk masses become comparable to planetary masses \citep[e.g.,][]{Matsumura09}. 
Since the sub-Neptune-size planets are more strongly affected by a residual gas disk, we may be overestimating the eccentricity excitation rate, if the instability occurs while the disk is still around.  
We evaluate the critical disk mass below which the eccentricity damping from the disk is negligible, by comparing the secular timescale with the eccentricity damping timescale $\tau_{sec} \lesssim \tau_{edamp}$. The critical surface mass density $\Sigma$ can be written as follows \citep[see Equations 45 and 49 in][]{Tanaka04}: 
%
\begin{equation}
\Sigma = \frac{1}{0.78\tau_{sec}}\left(\frac{M_p}{M_*}\right)^{-1} \left(\frac{c}{a\Omega_p}\right)^4 \frac{M_*}{a^2\Omega_p} \ ,
\end{equation}
%
where $M_p$ and $M_*$ are planetary and stellar masses, $c$ is the sound speed, and $\Omega_p$ is the orbital angular velocity of a planet.  By assuming the minimum mass solar nebula type disk with the surface mass density proportional to $a^{-3/2}$ as well as the temperature profile of $T=280(a/AU)^{-1/2}$, and by integrating the critical surface mass density from 0.01 to 100\,AU, we estimate the critical total disk mass that is required to damp a test particle's eccentricity at each radius (see the top panel of Figure~\ref{tsec_tedamp_sigma}). Here, we approximate the secular timescale as $\tau_{sec}\sim 2\pi/A$, by assuming the initial distribution of our default simulations.  In the bottom panel, we compare the timescale associated with a free frequency of a test particle $A$ with that associated with eigen frequencies of giants $g_i$. At the secular resonance locations $A\sim g_i$, the eccentricities increase most dramatically.   
The figure indicates that about a Jupiter-mass disk is required to prevent the secular eccentricity excitation of test particles due to giant planets.  Since such a mass is comparable to the critical disk mass for the dynamical instability of giant planets, the eccentricity damping effect from the disk may not be significant even for sub-Neptune-size planets.  Moreover, this is also comparable to the disk's mass where photoevaporation effects become significant \citep[e.g.,][]{Matsuyama03,Alexander06b}.  Therefore, even if a disk damps the eccentricity of sub-Neptune-size planets, that may be a short-lived phenomenon.  
These small-mass planets could also drift toward the central star due to the gravitational interactions with the disk.
Since this type I migration timescale is a factor of $(a\Omega_p/c)^2=(a/h)^2\sim1000$ longer than the eccentricity damping time $\tau_{edamp}$ ($h$ is a pressure scale height of the disk), such a migration would be negligible toward the end of the disk's lifetime.  
If the disk's effect turns out to be significant, sub-Neptune-size planets may safely survive the era of dynamical instability among giant planets.  In such a case, the actual survival region of these planets is likely to be somewhere between what we have statistically predicted here from the dynamical evolution, and what previous studies estimated from the current distributions of planets.  
  
The gas disks also provide additional precessions of planetary orbits and the dissipation of gas disks can move the locations of secular resonances \citep{Nagasawa00}.  
Such an effect can influence the overall results, but is beyond the scope of this paper. 
The other effects which precess the planetary orbits and change the resonance locations include general relativistic corrections as well as 
rotational and tidal oblateness effects of a star and a planet.  
However, these effects are important only near the central stars, and thus are unlikely to change our conclusions drastically.

\begin{figure}
\plotone{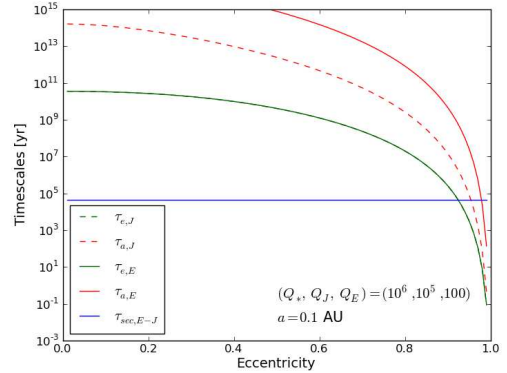}
\caption{Comparison of various timescales.  
Blue solid line shows the secular timescale between an Earth-like planet at
$0.1\,$AU and a Jupiter-like planet at $1\,$AU.  
Solid and dashed green curves show tidal circularization timescales for an
Earth-like and a Jupiter-like planet at $0.1\,$AU by assuming tidal quality
factors of $100$ and $10^5$, respectively.       
Solid and dashed red curves are corresponding orbital decay timescales for an
Earth-like and a Jupiter-like planet, respectively.
Secular timescale becomes less than circularization timescale for an Earth-like
planet for an eccentricity $e>0.9$.  From the orbital decay timescale, we find
that such a circularized Earth-like planet could survive for up to $0.1-1\,$Gyr.
\label{taua_taue_a0.1AU}}
\end{figure}
Our results have some theoretical predictions and observational implications as well:
\begin{itemize}
\item{{\it Formation of very close-in sub-Neptunes by scattering.}
We have not taken account of the tidal interactions between planets and the central star.  
With tidal effects, the highly eccentric giant planets can become close-in planets due to orbital decay and circularization \citep[e.g.][]{Fabrycky07,Wu07,Nagasawa08,Naoz11,Wu11}. Such a giant planet may remove closer-in terrestrial planets which survived in our simulations.    
On the other hand, highly eccentric sub-Neptune-mass planets could also achieve very short period, circular orbits, if orbital circularization timescales are short compared to secular timescales.  
We show comparisons of some relevant timescales in Figure~\ref{taua_taue_a0.1AU}.  
For an Earth-like planet at $0.1\,$AU which is accompanied by a Jupiter-like planet at $1\,$AU, the secular timescale $\tau_{sec,\,E-J}$ becomes shorter than the tidal circularization timescale $\tau_{e,\,E}$ for an eccentricity $e>0.9$.  
The corresponding orbital decay timescale $\tau_{a,\,E}$ for such a case can be up to $0.1-1\,$Gyr, 
which indicates that such a planet could survive for a typical age of a planet-hosting star.  
Similar tidal dissipation timescales for a Jupiter-like planet at $0.1\,$AU (not $1\,$AU) are also plotted in dashed curves for comparison.  Note that the circularization timescale for a Jupiter-like planet becomes comparable to an Earth-like planet for tidal quality factors of $Q=10^5$ and $100$, respectively. \cite{Raymond08a} proposed many scenarios to produce hot rocky planets, including what we suggested here.}

\item{{\it Metallicity enhancement of the stellar surface due to the mergers.}
Since our study suggests that many terrestrial planets and/or planetary embryos could merge with the central star under violent evolution of giant planets, 
it is informative to estimate whether such mergers would change the stellar metallicity significantly.
The Sun-like stars are fully convective during the pre-main-sequence stage, and thus the stellar metallicities would not be affected by merging planets which migrated through the gas disks.
However, the dynamical mechanism described here would occur toward the end of the disk's lifetime or later ($\gtrsim$ a few Myr), which is comparable to the timescale of the shrinkage of the convective region of a star.  
Thus, mergers of planets with a star in this stage could have a higher impact on the stellar metallicity than those during the earlier stage.  
For typical Sun-like stars in the main-sequence stage, a convective envelope accounts for a few percent of the total mass (about $0.02\,M_{\odot} \sim 4\times10^{28}\,$kg), and about $2\,\%$ of this mass is expected to be composed of metals ($\sim 8\times10^{26}\,$kg).  
When we write the stellar metallicity as 
$[{\rm Fe/H}] = \log[Z/X]-\log[Z_{\odot}/X_{\odot}]$, 
and assume that the mergers of terrestrial planets only change the mass fraction of 
metals Z, 
the total mass of $\sim35\,M_E$ is required to change the stellar metallicity by $0.1$ from the solar value.  
Thus, a significant change of a stellar metallicity might be possible if there were multiple mergers of Super Earths. 
However, it could be difficult to tell the primordial metallicity from such a raised value by mergers.}  

\item{{\it Lack of sub-Neptunes around metal-rich stars.}
The giant planets have been found around solar-type stars with preferences for metal-rich stars.  
On the other hand, sub-Neptune-like planets are not found around metal-rich stars with $[{\rm Fe/H}]\gtrsim0.2$ \citep{Mayor11ap}. 
Since higher metallicities imply higher formation rates of giant planets, it is possible that sub-Neptune-mass planets in metal-rich systems tend to be exposed to dynamical instabilities of giant planets.
Then, dynamical instabilities would remove these sub-Neptune-mass planets as described in this paper, and they could be rarer among metal-rich systems.
Alternatively, a recent Monte Carlo population synthesis study by \cite{Mordasini12ap} showed that their single planet formation model produced more sub-Neptune-mass planets around lower-metallicity stars. 
Currently, the metallicity distribution of Kepler's planet-hosting stars is poorly known, though there is an indication that these stars are metal rich compared to stars with no planets \citep{Schlaufman11}. 
If the present metallicity trend were confirmed by future observations including Kepler, HARPS, and ESPRESSO, we would need to understand whether such a trend is due to natural planet formation or evolution involving dynamically unstable giant planets. } 

\end{itemize}
%
%
%
%
%
\acknowledgements{We thank Makoto Obinata who carried out the preliminary study.  
We also thank two referees who submitted the reviews promptly after the originally assigned referee withdrew.  
SM is supported by an Astronomy Center for Theory and Computation Prize Fellowship at the University of Maryland. 
NM is supported by JSPS KAKENHI (21740324).}
%
%
%
%
%
%

\bibliographystyle{apj}
\bibliography{REF}

\begin{thebibliography}{48}
\expandafter\ifx\csname natexlab\endcsname\relax\def\natexlab#1{#1}\fi

\bibitem[{{Adams} \& {Laughlin}(2003)}]{Adams03}
{Adams}, F.~C., \& {Laughlin}, G. 2003, Icarus, 163, 290

\bibitem[{{Alexander} {et~al.}(2006){Alexander}, {Clarke}, \&
  {Pringle}}]{Alexander06b}
{Alexander}, R.~D., {Clarke}, C.~J., \& {Pringle}, J.~E. 2006, MNRAS, 369, 229

\bibitem[{{Armitage}(2003)}]{Armitage03}
{Armitage}, P.~J. 2003, ApJL, 582, L47

\bibitem[{{Asghari} {et~al.}(2004){Asghari}, {Broeg}, {Carone},
  {Casas-Miranda}, {Castro Palacio}, {Csillik}, {Dvorak}, {Freistetter},
  {Hadjivantsides}, {Hussmann}, {Khramova}, {Khristoforova}, {Khromova},
  {Kitiashivilli}, {Kozlowski}, {Laakso}, {Laczkowski}, {Lytvinenko}, {Miloni},
  {Morishima}, {Moro-Martin}, {Paksyutov}, {Pal}, {Patidar}, {Pe{\v c}nik},
  {Peles}, {Pyo}, {Quinn}, {Rodriguez}, {Romano}, {Saikia}, {Stadel}, {Thiel},
  {Todorovic}, {Veras}, {Vieira Neto}, {Vilagi}, {von Bloh}, {Zechner}, \&
  {Zhuchkova}}]{Asghari04}
{Asghari}, N., {Broeg}, C., {Carone}, L., {et~al.} 2004, A\&A, 426, 353

\bibitem[{{Batalha} {et~al.}(2012){Batalha}, {Rowe}, {Bryson}, {Barclay},
  {Burke}, {Caldwell}, {Christiansen}, {Mullally}, {Thompson}, {Brown},
  {Dupree}, {Fabrycky}, {Ford}, {Fortney}, {Gilliland}, {Isaacson}, {Latham},
  {Marcy}, {Quinn}, {Ragozzine}, {Shporer}, {Borucki}, {Ciardi}, {Gautier},
  {Haas}, {Jenkins}, {Koch}, {Lissauer}, {Rapin}, {Basri}, {Boss}, {Buchhave},
  {Charbonneau}, {Christensen-Dalsgaard}, {Clarke}, {Cochran}, {Demory},
  {Devore}, {Esquerdo}, {Everett}, {Fressin}, {Geary}, {Girouard}, {Gould},
  {Hall}, {Holman}, {Howard}, {Howell}, {Ibrahim}, {Kinemuchi}, {Kjeldsen},
  {Klaus}, {Li}, {Lucas}, {Morris}, {Prsa}, {Quintana}, {Sanderfer},
  {Sasselov}, {Seader}, {Smith}, {Steffen}, {Still}, {Stumpe}, {Tarter},
  {Tenenbaum}, {Torres}, {Twicken}, {Uddin}, {Van Cleve}, {Walkowicz}, \&
  {Welsh}}]{Batalha12ap}
{Batalha}, N.~M., {Rowe}, J.~F., {Bryson}, S.~T., {et~al.} 2012, ArXiv e-prints

\bibitem[{{Chambers}(1999)}]{Chambers99}
{Chambers}, J.~E. 1999, MNRAS, 304, 793

\bibitem[{{Chambers} {et~al.}(1996){Chambers}, {Wetherill}, \&
  {Boss}}]{Chambers96}
{Chambers}, J.~E., {Wetherill}, G.~W., \& {Boss}, A.~P. 1996, Icarus, 119, 261

\bibitem[{{Chatterjee} {et~al.}(2008){Chatterjee}, {Ford}, {Matsumura}, \&
  {Rasio}}]{Chatterjee08}
{Chatterjee}, S., {Ford}, E.~B., {Matsumura}, S., \& {Rasio}, F.~A. 2008, ApJ,
  686, 580

\bibitem[{{Edgar} \& {Artymowicz}(2004)}]{Edgar04}
{Edgar}, R., \& {Artymowicz}, P. 2004, MNRAS, 354, 769

\bibitem[{{Fabrycky} \& {Tremaine}(2007)}]{Fabrycky07}
{Fabrycky}, D., \& {Tremaine}, S. 2007, ApJ, 669, 1298

\bibitem[{{Fogg} \& {Nelson}(2005)}]{Fogg05}
{Fogg}, M.~J., \& {Nelson}, R.~P. 2005, A\&A, 441, 791

\bibitem[{{Ford} \& {Rasio}(2008)}]{Ford08}
{Ford}, E.~B., \& {Rasio}, F.~A. 2008, ApJ, 686, 621

\bibitem[{{Gehman} {et~al.}(1996){Gehman}, {Adams}, \& {Laughlin}}]{Gehman96}
{Gehman}, C.~S., {Adams}, F.~C., \& {Laughlin}, G. 1996, PASP, 108, 1018

\bibitem[{{Gladman}(1993)}]{Gladman93}
{Gladman}, B. 1993, Icarus, 106, 247

\bibitem[{{Ito} \& {Tanikawa}(2001)}]{Ito01}
{Ito}, T., \& {Tanikawa}, K. 2001, PASJ, 53, 143

\bibitem[{{Jones} {et~al.}(2001){Jones}, {Sleep}, \& {Chambers}}]{Jones01}
{Jones}, B.~W., {Sleep}, P.~N., \& {Chambers}, J.~E. 2001, A\&A, 366, 254

\bibitem[{{Juri{\'c}} \& {Tremaine}(2008)}]{Juric08}
{Juri{\'c}}, M., \& {Tremaine}, S. 2008, ApJ, 686, 603

\bibitem[{{Levison} \& {Agnor}(2003)}]{Levison03}
{Levison}, H.~F., \& {Agnor}, C. 2003, AJ, 125, 2692

\bibitem[{{Lin} {et~al.}(1996){Lin}, {Bodenheimer}, \& {Richardson}}]{Lin96}
{Lin}, D.~N.~C., {Bodenheimer}, P., \& {Richardson}, D.~C. 1996, Nature, 380,
  606

\bibitem[{{Lufkin} {et~al.}(2006){Lufkin}, {Richardson}, \& {Mundy}}]{Lufkin06}
{Lufkin}, G., {Richardson}, D.~C., \& {Mundy}, L.~G. 2006, ApJ, 653, 1464

\bibitem[{{Mandell} \& {Sigurdsson}(2003)}]{Mandell03}
{Mandell}, A.~M., \& {Sigurdsson}, S. 2003, ApJL, 599, L111

\bibitem[{{Marchal} \& {Bozis}(1982)}]{Marchal82}
{Marchal}, C., \& {Bozis}, G. 1982, Celestial Mechanics, 26, 311

\bibitem[{{Matsumura} {et~al.}(2009){Matsumura}, {Pudritz}, \&
  {Thommes}}]{Matsumura09}
{Matsumura}, S., {Pudritz}, R.~E., \& {Thommes}, E.~W. 2009, ApJ, 691, 1764

\bibitem[{{Matsuyama} {et~al.}(2003){Matsuyama}, {Johnstone}, \&
  {Hartmann}}]{Matsuyama03}
{Matsuyama}, I., {Johnstone}, D., \& {Hartmann}, L. 2003, ApJ, 582, 893

\bibitem[{{Mayor} {et~al.}(2011){Mayor}, {Marmier}, {Lovis}, {Udry},
  {S{\'e}gransan}, {Pepe}, {Benz}, {Bertaux}, {Bouchy}, {Dumusque}, {Lo Curto},
  {Mordasini}, {Queloz}, \& {Santos}}]{Mayor11ap}
{Mayor}, M., {Marmier}, M., {Lovis}, C., {et~al.} 2011, ArXiv e-prints

\bibitem[{{Menou} \& {Tabachnik}(2003)}]{Menou03}
{Menou}, K., \& {Tabachnik}, S. 2003, ApJ, 583, 473

\bibitem[{{Moorhead} \& {Adams}(2008)}]{Moorhead08}
{Moorhead}, A.~V., \& {Adams}, F.~C. 2008, Icarus, 193, 475

\bibitem[{{Mordasini} {et~al.}(2012){Mordasini}, {Alibert}, {Benz}, {Klahr}, \&
  {Henning}}]{Mordasini12ap}
{Mordasini}, C., {Alibert}, Y., {Benz}, W., {Klahr}, H., \& {Henning}, T. 2012,
  ArXiv e-prints

\bibitem[{{Murray} \& {Dermott}(1999)}]{Murray99}
{Murray}, C.~D., \& {Dermott}, S.~F. 1999, {Solar system dynamics} (Solar
  system dynamics by Murray, C.~D., 1999)

\bibitem[{{Nagasawa} {et~al.}(2008){Nagasawa}, {Ida}, \& {Bessho}}]{Nagasawa08}
{Nagasawa}, M., {Ida}, S., \& {Bessho}, T. 2008, ApJ, 678, 498

\bibitem[{{Nagasawa} {et~al.}(2000){Nagasawa}, {Tanaka}, \& {Ida}}]{Nagasawa00}
{Nagasawa}, M., {Tanaka}, H., \& {Ida}, S. 2000, AJ, 119, 1480

\bibitem[{{Naoz} {et~al.}(2011){Naoz}, {Farr}, {Lithwick}, {Rasio}, \&
  {Teyssandier}}]{Naoz11}
{Naoz}, S., {Farr}, W.~M., {Lithwick}, Y., {Rasio}, F.~A., \& {Teyssandier}, J.
  2011, Nature, 473, 187

\bibitem[{{Noble} {et~al.}(2002){Noble}, {Musielak}, \& {Cuntz}}]{Noble02}
{Noble}, M., {Musielak}, Z.~E., \& {Cuntz}, M. 2002, ApJ, 572, 1024

\bibitem[{{Raymond} {et~al.}(2008){Raymond}, {Barnes}, \&
  {Mandell}}]{Raymond08a}
{Raymond}, S.~N., {Barnes}, R., \& {Mandell}, A.~M. 2008, MNRAS, 384, 663

\bibitem[{{Raymond} {et~al.}(2006){Raymond}, {Mandell}, \&
  {Sigurdsson}}]{Raymond06}
{Raymond}, S.~N., {Mandell}, A.~M., \& {Sigurdsson}, S. 2006, Science, 313,
  1413

\bibitem[{{Raymond} {et~al.}(2011){Raymond}, {Armitage}, {Moro-Mart{\'{\i}}n},
  {Booth}, {Wyatt}, {Armstrong}, {Mandell}, {Selsis}, \& {West}}]{Raymond11}
{Raymond}, S.~N., {Armitage}, P.~J., {Moro-Mart{\'{\i}}n}, A., {et~al.} 2011,
  A\&A, 530, A62

\bibitem[{{Raymond} {et~al.}(2012){Raymond}, {Armitage}, {Moro-Mart{\'{\i}}n},
  {Booth}, {Wyatt}, {Armstrong}, {Mandell}, {Selsis}, \& {West}}]{Raymond12}
---. 2012, A\&A, 541, A11

\bibitem[{{Schlaufman} \& {Laughlin}(2011)}]{Schlaufman11}
{Schlaufman}, K.~C., \& {Laughlin}, G. 2011, ApJ, 738, 177

\bibitem[{{Tanaka} \& {Ida}(1999)}]{Tanaka99}
{Tanaka}, H., \& {Ida}, S. 1999, Icarus, 139, 350

\bibitem[{{Tanaka} \& {Ward}(2004)}]{Tanaka04}
{Tanaka}, H., \& {Ward}, W.~R. 2004, ApJ, 602, 388

\bibitem[{{Th{\'e}bault} {et~al.}(2002){Th{\'e}bault}, {Marzari}, \&
  {Scholl}}]{Thebault02}
{Th{\'e}bault}, P., {Marzari}, F., \& {Scholl}, H. 2002, A\&A, 384, 594

\bibitem[{{Veras} \& {Armitage}(2005)}]{Veras05}
{Veras}, D., \& {Armitage}, P.~J. 2005, ApJL, 620, L111

\bibitem[{{Veras} \& {Armitage}(2006)}]{Veras06}
---. 2006, ApJ, 645, 1509

\bibitem[{{Ward} \& {Hahn}(1995)}]{Ward95}
{Ward}, W.~R., \& {Hahn}, J.~M. 1995, ApJL, 440, L25

\bibitem[{{Wright} {et~al.}(2009){Wright}, {Upadhyay}, {Marcy}, {Fischer},
  {Ford}, \& {Johnson}}]{Wright09}
{Wright}, J.~T., {Upadhyay}, S., {Marcy}, G.~W., {et~al.} 2009, ApJ, 693, 1084

\bibitem[{{Wu} \& {Lithwick}(2011)}]{Wu11}
{Wu}, Y., \& {Lithwick}, Y. 2011, ApJ, 735, 109

\bibitem[{{Wu} {et~al.}(2007){Wu}, {Murray}, \& {Ramsahai}}]{Wu07}
{Wu}, Y., {Murray}, N.~W., \& {Ramsahai}, J.~M. 2007, ApJ, 670, 820

\bibitem[{{Zhou} {et~al.}(2005){Zhou}, {Aarseth}, {Lin}, \&
  {Nagasawa}}]{Zhou05}
{Zhou}, J.-L., {Aarseth}, S.~J., {Lin}, D.~N.~C., \& {Nagasawa}, M. 2005, ApJL,
  631, L85

\end{thebibliography}

\end{document}